\documentclass[aps,prfluids,reprint,onecolumn,linenumbers,superscriptaddress]{revtex4-2}
\usepackage{natbib}
\usepackage{graphicx}
\usepackage{dcolumn}
\usepackage{bm}
\usepackage{amsmath}
\usepackage{xcolor}
\usepackage{hyperref}

\let\oldequation\equation
\let\oldendequation\endequation
\renewenvironment{equation}{\linenomathNonumbers\oldequation}{\oldendequation\endlinenomath}
\let\oldgather\gather
\let\oldendgather\endgather
\renewenvironment{gather}{\linenomathNonumbers\oldgather}{\oldendgather\endlinenomath}

\renewcommand*{\v}[1]{\boldsymbol{#1}}
\newcommand*{\vv}[1]{\mathsf{#1}}
\newcommand*{\subtext}[1]{\mathrm{#1}}
\newcommand*{\grad}{\boldsymbol{\nabla}}
\providecommand*{\bcdot}{\boldsymbol{\cdot}}
\renewcommand*{\div}{\grad\bcdot}
\newcommand*{\Tr}{\mathop{\mathrm{Tr}}}
\newcommand*{\Rout}{R_{\subtext{out}}}
\newcommand*{\Ca}{Ca} 
\newcommand*{\pd}{\partial}
\newcommand*{\asp}{\Rout}
\newcommand*{\fsq}{Q_0}
\newcommand*{\fst}{\Gamma}
\newcommand*{\fca}{\mathcal{C}}
\newcommand*{\fht}{\mathcal{H}}
\newcommand*{\Rinit}{R_{\subtext{init}}}
\newcommand*{\Vinit}{V_{\subtext{init}}}
\newcommand*{\bmin}{b_{\subtext{min}}}
\newcommand*{\ellrim}{\ell_{\subtext{rim}}}
\newcommand*{\front}{{\subtext{front}}}
\newcommand*{\bulge}{{\subtext{bulge}}}
\newcommand*{\transpose}{\ensuremath{T}} 

\newcommand*{\figrefo}[1]{Fig.~\ref{#1}} 
\newcommand*{\Figrefo}[1]{Fig.~\ref{#1}} 
\newcommand*{\appref}[1]{Appendix~\ref{#1}}
\newcommand*{\secref}[1]{Sec.~\ref{#1}}

\newcommand*{\figparen}[1]{[#1]} 
\newcommand*{\figa}[1]{(#1)} 
\newcommand*{\figref}[2][\relax]{\ifx\relax#1\figrefo{#2}\else\figrefo{#2}\figa{#1}\fi} 
\newcommand*{\Figref}[2][\relax]{\ifx\relax#1\Figrefo{#2}\else\Figrefo{#2}\figa{#1}\fi} 
\newcommand*{\figrefp}[2][\relax]{\ifx\relax#1(\figrefo{#2})\else\figparen{\figrefo{#2}\figa{#1}}\fi} 
\newcommand*{\Figrefp}[2][\relax]{\ifx\relax#1(\Figrefo{#2})\else\figparen{\Figrefo{#2}\figa{#1}}\fi} 

\newcommand*{\eqparen}[1]{[#1]} 
\newcommand*{\Eqreft}[1]{Eq.~\eqref{#1}}
\newcommand*{\eqreft}[1]{Eq.~\eqref{#1}}
\newcommand*{\eqrefp}[1]{[Eq.~\eqref{#1}]}

\newcommand*{\eqrefsp}[1]{[Eqs.~\eqref{#1}]}
\newcommand*{\eqrefsst}[3]{Eqs.~\eqref{#1}#2\eqref{#3}}

\begin{document}
\nolinenumbers

\title{Axisymmetric gas--liquid displacement flow under a confined elastic slab}

\author{Gunnar G. Peng}
\affiliation{Department of Physics \& Astronomy and Manchester Centre for Nonlinear Dynamics, University of Manchester, Oxford Road, M13 9PL, Manchester, UK}
\affiliation{Department of Mathematics, Imperial College London, London SW7 2AZ, UK}
\author{Callum Cuttle}
\affiliation{Department of Engineering Science, University of Oxford, Parks Road, Oxford, OX1 3PJ, UK}
\author{Christopher W. MacMinn}
\affiliation{Department of Engineering Science, University of Oxford, Parks Road, Oxford, OX1 3PJ, UK}
\author{Draga Pihler-Puzovi\'{c}}
\affiliation{Department of Physics \& Astronomy and Manchester Centre for Nonlinear Dynamics, University of Manchester, Oxford Road, M13 9PL, Manchester, UK}

\date{\today}

\begin{abstract}
A circular Hele-Shaw cell bounded by a volumetrically confined elastic solid can act as a fluidic fuse: during radially outward fluid flow, the solid deforms in response to the viscous pressure field such that the gap expands near the inlet (at the centre) and contracts near the outlet (around the rim). If the flow rate exceeds a critical value, the gap at the outlet can close completely, interrupting/choking the flow. Here, we consider the injection of gas into such a soft-walled Hele-Shaw cell filled with viscous liquid. Our theoretical model and numerical simulations for axisymmetric flow driven by the injection of an expanding gas bubble show that the bubble increases the critical flow rate of choking via two mechanisms. Firstly, as the interface approaches the rim, it reduces the length over which the viscous pressure gradient deforms the solid, which increases the critical flow rate above which choking occurs. Secondly, compression of the gas reduces the outlet flow rate relative to the inlet flow rate. As a consequence, for large injection rates, a near-choking regime is established in which the outlet flow rate becomes independent of the injection rate and instead depends only on the instantaneous position of the interface. Our travelling-wave model for the advancement of the bubble front will enable future reduced-order modelling of non-axisymmetric problems, such as viscous fingering.
\end{abstract}

\maketitle

\section{Introduction}

From flow through porous media \citep{Lee2020} to passive microfluidics \citep{Stone2009}, interaction of two-phase viscous flows with soft deformable components (low-Re FSI)  is a staple of many natural and industrial settings. For example, in the pulmonary airway tree under pathological conditions, air entering the compliant lungs encounters plugs of mucus that occlude its passageways \citep{Heil2015}. Other types of low-Re FSI have recently been exploited for technological progress, for example, in improving the manufacturing quality of products \citep{Chong2007}, personalising diagnostic tools \citep{Lin2020} and developing soft robotics \citep{Jones2021}. Fundamental understanding of such complex flows can be gained by studying much simpler model problems.

In this paper, we study gas--liquid displacement in a soft Hele-Shaw cell. Its rigid counterpart, which consists of a narrow gap between two parallel plates, is a classical model system for studying fluid--fluid displacement. At low flow rates, the interface between the two fluids is always circular (i.e. stable). However, if a less viscous fluid (e.g.~an air bubble) displaces a more viscous fluid (e.g.~glycerol) at a sufficiently high rate,  the interface becomes unstable and develops distinct fingers that subsequently compete, split, and branch, forming a complex interfacial pattern \citep{Saffman58}. This fingering instability can be suppressed to higher flow rates if one of the walls of the Hele-Shaw cell is replaced by a thin, unconfined elastic membrane, which allows the injected volume to be accommodated in large part by inflation rather than viscous displacement \citep{PihlerPuzovic12, Juel18}. Interestingly, the deformation of the flow cell remains roughly axisymmetric and independent of the morphology of the displacement front~\citep{PihlerPuzovic2018}, unless the elastic membrane is very compliant \citep{PihlerPuzovic2014}.

\begin{figure}
\includegraphics[width=\linewidth]{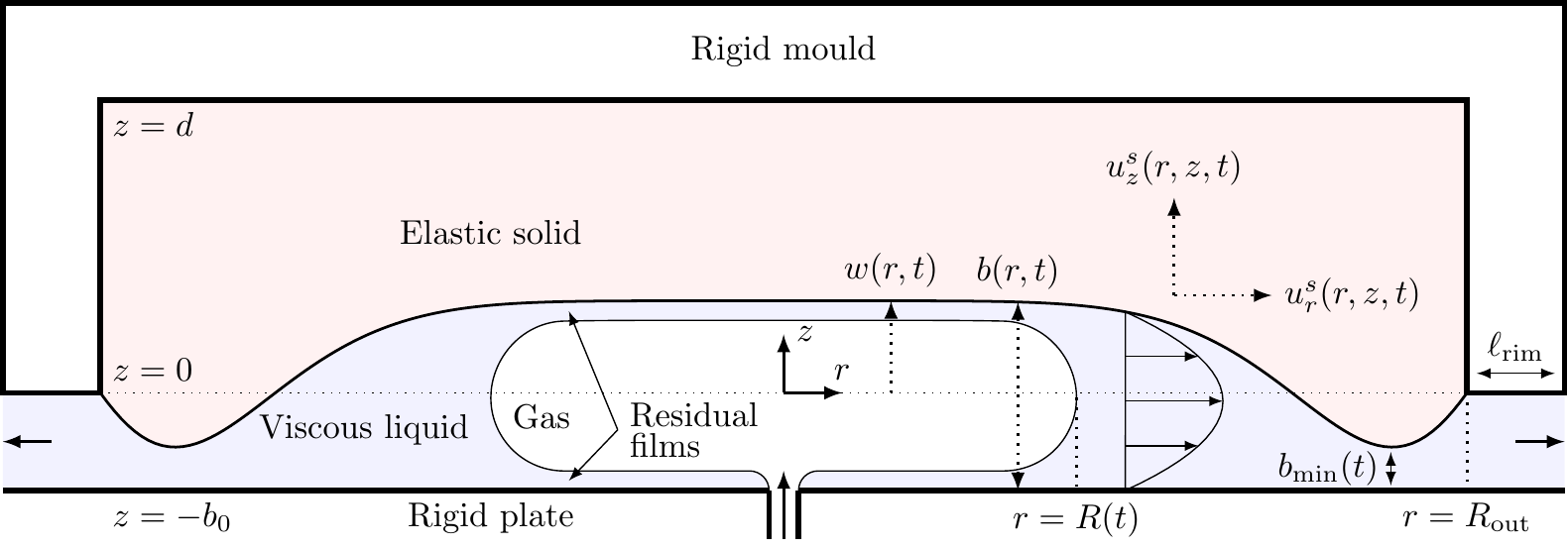}
\caption{Schematic of radially outward gas--liquid displacement in a Hele-Shaw cell bounded by a confined elastic solid. \label{fig:setup}}
\end{figure}

\begin{figure}
\includegraphics[width=\linewidth]{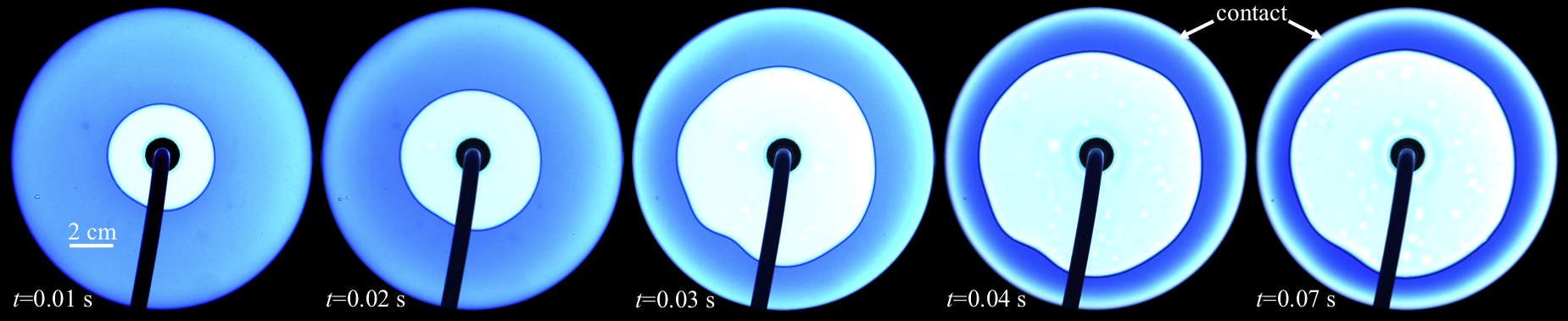}
\caption{Top view images from preliminary experiments with an approximately circular interface at various times $t$ from the start of the experiment. The injected gas bubble displaces glycerol (dyed blue) in the narrow gap of a soft Hele-Shaw cell.
Flow-induced deformation of the elastomer eventually leads to contact between the soft slab and the glass plate in the vicinity of the cell rim (visible as a white band encircling the cell), trapping the interface within the cell. The experimental parameters are: cell radius $\Rout = 60$~mm and initial gap $b_0 = 2$~mm; elastomer thickness $d = 15$~mm, shear modulus $G = 1.36$~kPa and Poisson's ratio $\nu = 0.495$; liquid viscosity $\mu=0.72$~Pa.s and surface tension $\gamma = 63$~mN/m, imposed flow rate $Q_0= 450$~ml/min.\label{fig:exp}}
\end{figure}

The soft Hele-Shaw cells considered here contain a volumetrically confined slab of elastomer, as shown schematically in \figref{fig:setup}. Fluid--structure interaction (FSI) between a viscous pressure gradient and the confined elastomer in such cells leads to the accumulation of soft material near the cell rim, which constricts the gap \citep{Box2020, Peng2022}. The elastomer can even make contact with the opposite rigid wall and choke the flow entirely, as in \figref{fig:exp}. In the absence of a gas-liquid interface, choking occurs systematically above a critical injection flow rate \citep{Box2020}. However, injection of gas into the soft cell brings a number of key differences. Firstly, gas is compressible, so the rate of change of gas volume in the cell varies in time. Secondly, its viscosity is negligible, so the region over which the viscous pressure gradient acts reduces over time as the liquid gets displaced. Capillarity is also important at the interface between two fluids, and, as discussed above, the interface is prone to viscous fingering, though choking has been observed even when the interface remained approximately axisymmetric throughout the experiment~\figrefp{fig:exp}. The role of viscous fingering onto choking has been studied recently in gas--liquid displacement experiments by~\citet{Peng2022}. Unlike in the inflatable Hele-Shaw cells, non-axisymmetry of the fingered interface in the soft cell studied here directly affects the deformation of the confined soft wall, which results in a complex choking threshold. We depart from the previous study of \citet{Peng2022}, and focus instead on the role of gas compressibility and viscous pressure gradients on choking in an axisymmetric geometry by exploring a mathematical model of an axisymmetric two-phase lubrication flow under a confined elastic slab. Thus, we decouple the influence of the two-phase displacement from that of the viscous fingering in experiments by \citet{Peng2022}.
  
In existing literature, gas compression is often neglected or carefully avoided, e.g.~by extracting liquids rather than injecting gas \citep{Park1984}. However, compressibility effects are unavoidable in many practical settings, e.g. during the gas-driven displacement of granular suspensions \citep{Sandnes11}, during gas invasion into liquid-saturated porous media \citep{Lee19}, during foam-driven hydraulic fracturing \citep{Lai2018}, or in soft microfluidics carrying a viscous flow with a small amount of air trapped in the system \citep{Guyard2022}. Few studies have addressed the role of gas compressibility in the dynamics of a gas-liquid displacement flow, but it is known to be strongly coupled to viscous displacement in, e.g., frictional flows \citep{Sandnes11}, capillary tubes \citep{Cuttle2021} and rigid Hele-Shaw cells \citep{Cuttle2022}. Here we go a step further and additionally consider the coupling of gas compression with both liquid displacement and elastic deformation of the flow cell. 

This paper is laid out as follows. We present the axisymmetric governing equations and non-dimensionalisation in \secref{sec:model}, followed by a description of the gas--liquid displacement flow at low injection rates in \secref{sec:num}. For higher flow rates, we study the impact of gas--liquid displacement on choking in \secref{sec:choke}, by relating the deformation of the elastomer to the proximity of the bubble to the cell rim (\secref{sec:proximity}) and identifying a near-choking regime when the compressibility of the gas is significant (\secref{sec:nearchoke}). We investigate the dynamics of the advancing bubble front in \secref{sec:tw}. We summarise and discuss the results in \secref{sec:summ}.

\section{Theoretical model \label{sec:model}}

\subsection{Governing equations \label{subsec:gov}}

The setup is shown in \figref{fig:setup}. We consider a Hele-Shaw cell of initial (relaxed) gap thickness $b_0$, bounded by a rigid wall below and by an elastic slab above. The elastic slab is a cylinder of radius $\Rout$, thickness $d \lesssim \Rout$, and shear modulus $G$ that is confined both around the outer rim and from above within a rigid mould. The cell is initially filled with liquid of viscosity $\mu$. A gas bubble is injected at the nominal volumetric flow rate $Q_0$ at the centre of the cell, displacing the liquid and also deforming the elastomer. We neglect inertia and gravity, as well as the compressibility of the solid and of the liquid.

We employ cylindrical polar coordinates $(r,\theta,z)$ with the surface of the undeformed elastic solid located at $z=0$, and the centre of the cell at $r=0$. We assume axisymmetry, as discussed above, so that there is no explicit dependence on the azimuthal angle $\theta$. 

One key assumption in our analysis is that the initial cell gap $b_0$ and the vertical deformation $w$ are small compared with the initial slab thickness $d$, i.e.\ that $b_0,w \ll d$. As a result, the gap thickness can change significantly from its initial value, while the strains in the elastic solid remain small, allowing us to adopt linear elasticity. For a deformation characterised by  displacement $\v{u}^s$, stress tensor $\vv\sigma^s$ and pressure $p^s = -(\Tr \vv\sigma^s)/3$, the equations for linear elasticity, incompressibility and mechanical equilibrium in the solid take the form
\begin{subequations}\label{eq:elastic}
\begin{gather}
	\vv\sigma^s = -p^s\vv{I} + G(\grad \v{u}^s + (\grad\v{u}^s)^\transpose), \qquad \div \v{u}^s = 0, 
	\\
	\v{0} = \div\vv\sigma^s = -\grad p^s + G\nabla^2 \v{u}^s,
\end{gather}
\end{subequations}
in the solid domain $0 \leq r \leq \Rout$, $0 \leq z \leq d$. Here, $\Tr$ denotes the trace, $\grad = \v{e}_r \pd_r + \v{e}_\theta (1/r) \pd_\theta + \v{e}_z \pd_z$ is the gradient operator with $\v{e}_r$, $\v{e}_\theta$ and $\v{e}_z$ the coordinate unit vectors, $\vv{I}$ is the identity tensor, and superscript $\transpose$ denotes transpose. We impose that the solid is adhered to the mould and that there is no singularity at the centre,
\begin{equation}
  \v{u}^s = \v{0} \text{ at } r = \Rout \text{ and at } z = d, \qquad u_r^s = \pd_r u_z^s = 0 \text{ at } r = 0.
\end{equation}
The solid is coupled to the flow in the gap via the vertical displacement $w(r,t)$ of the surface and the gauge pressure $p(r,t)$ on the surface (measured relative to atmospheric pressure), while the viscous shear stress from the flow on the surface can be neglected due to the assumption that $b_0,w \ll d$,
\begin{equation}
	u_z^s|_{z=0} = w, \qquad \sigma_{zz}^s|_{z=0} = -p, \qquad \sigma_{rz}^s|_{z=0} = 0.
\end{equation}

The local gap $b(r,t)$ is related to the vertical deformation $w(r,t)$ of the solid surface by
\begin{equation}
	b(r,t) = b_0+w(r,t).
\end{equation}
For the flow, we split the domain into two parts. In the bubble region $r < R(t)$, the pressure is spatially uniform:
\begin{equation}\label{eq:bub}
	p(r,t) = p_b(t) \quad \text{ in } r < R(t). 
\end{equation}
In the liquid region $r > R(t)$, we adopt the standard Hele-Shaw assumption that the pressure is vertically uniform and equal to $p(r,t)$ to leading order in $b_0/\Rout$, satisfying the lubrication equation
\begin{equation}\label{eq:lub}
	\dot b = \grad_H \bcdot \left(\frac{b^3}{12\mu} \grad_H p\right) \quad \text{ in } r > R(t).
\end{equation}
Here, the over-dot is the partial derivative with respect to time and $\grad_H = \v{e}_r \pd_r + \v{e}_\theta (1/r)\pd_\theta$ is the horizontal gradient operator. (Note that we neglect any horizontal velocity from the solid onto the fluid, due to $b_0 \ll d$). We do not model the advancing gas--liquid interface at the displacement front $r = R(t)$ in detail. Instead, following \citet{Peng2015}, we employ approximate kinematic and dynamic boundary conditions appropriate for a growing bubble in a Hele-Shaw cell with rigid and parallel walls,
\begin{subequations}
\begin{equation}\label{eq:front}
	(1 - f_1)\dot R = -\frac{b^2}{12\mu}\frac{\pd p}{\pd r}, \qquad p - p_b = -\frac{2\gamma}{b}(1 + f_2) - \frac{\pi}{4} \frac{\gamma}{R} \qquad \text{ at } r = R^+.
\end{equation}
These conditions depend on the instantaneous capillary number $Ca = \mu \dot{R}/\gamma$ via two fitting functions,
\begin{equation}
	f_1(Ca)=\frac{Ca^{2/3}}{0.76+2.16\,Ca^{2/3}}, \qquad f_2(Ca)=\frac{Ca^{2/3}}{0.26+1.48\,Ca^{2/3}}+1.59\,Ca,
\end{equation}
\end{subequations}
which, respectively, describe the thickness of the residual liquid films on the cell walls behind the front and the additional pressure drop due to viscous resistance near the front.

Initially, the cell is undeformed and contains a small bubble of radius $\Rinit$ (which we take to be $\Rinit = d/2$ unless otherwise specified),
\begin{equation}
  w|_{t=0} = 0, \qquad R|_{t=0} = \Rinit.
\end{equation}
At the cell outlet $r = \Rout$, we impose atmospheric pressure (i.e.~zero gauge pressure) and let $Q(t)$ denote the flow rate of liquid leaving the cell, 
\begin{equation}\label{eq:injection}
	p|_{r=\Rout} = 0, \qquad Q(t) = -2\pi\Rout \left.\frac{b^3}{12\mu}\frac{\pd p}{\pd r}\right|_{r = \Rout}.
\end{equation}
Here, we have neglected the contribution to the viscous pressure drop from the thickness $\ellrim$ of the rim of the mould: Past the edge of the elastic solid, the rim creates a region of constant cell gap $b = b_0$, which could be accounted for by solving the lubrication equation \eqref{eq:lub} with the given flow rate $Q(t)$, resulting in the alternative pressure condition $p = (Q/2\pi)\ln (1 + \ellrim/\Rout)$ at $r = \Rout$, but we neglect this effect as $\ellrim/\Rout\ll1$, and use \eqreft{eq:injection} instead.

\subsection{Gas injection models} 

Due to incompressibility of the liquid and solid, the outlet flow rate $Q(t)$ is also the rate of change of gas volume in the cell. We assume in all cases that gas is injected at a constant nominal flow rate $Q_0$. If the compression of the gas is negligible, we simply have 
\begin{equation}\label{eq:flowrate_incomp} 
	Q(t) = Q_0.
\end{equation}
However, the elevated pressure $p_b(t)$ in the bubble compresses the gas, which may lead to a significant deviation between $Q(t)$ and $Q_0$. We assume that the heat generated by compression is rapidly lost to the environment, so that the gas can be approximated as isothermal. If the mass of gas in the system has volume $V_u(t)$ under atmospheric pressure $p_a$, then, after compression to an absolute pressure $p_a + p_b(t)$ its volume is $V_b = V_u/(1 + p_b/p_a)$. The compression of the injected gas proceeds differently depending on the method of its injection, and we consider two different methods that have been used in experiments \citep{Peng2022}. For injection using a syringe pump at a nominal rate $Q_0$, the pump chamber, tubing and bubble together form a sealed mass of gas with original volume $V_u = \Vinit$, so the flow rate is
\begin{subequations}\label{eq:flowrate_comp}
\begin{equation}\label{eq:flowrate_syringe}
	Q(t) = Q_0 + \dot V_b = Q_0 + \frac{\mathrm{d}}{\mathrm{d}t}\left[\frac{\Vinit}{(1 + p_b(t)/p_a)}\right] \qquad \text{(syringe pump)}.
\end{equation}
For injection from a pressurized gas bottle with pressure $\gg p_a$ via a needle resistor tuned to result in a fixed volume flow rate $Q_0$ of atmospheric-pressure gas downstream, the total uncompressed volume of air in the system increases as $V_u = \Vinit + Q_0 t$, where $\Vinit$ is the initial volume of air in the cell and the tubing downstream of the resistor, so the flow rate is 
\begin{equation}\label{eq:flowrate_bottle}
	Q(t) = \dot V_b = \frac{\mathrm{d}}{\mathrm{d}t} \left[\frac{\Vinit + Q_0 t}{(1 + p_b(t)/p_a)}\right] \qquad \text{(pressurised bottle)}.
\end{equation}
\end{subequations}

Although the two expressions \eqref{eq:flowrate_comp} are similar, and approximately equal when $\Vinit$ is sufficiently large \citep{Cuttle2021}, an important difference between the two injection methods is how small $\Vinit$ could reasonably be in practice. For injection using a syringe pump, the initial gas volume must be at least as large as the volume of the flow cell, so as to allow the injection to proceed until the bubble reaches the rim of the cell. For injection using a pressurised bottle, however, the initial gas volume can be much lower, just equal to the volume of the initial bubble in the cell, assuming that the tubing volume can be neglected. As we do not seek to investigate the effects of varying $\Vinit$ in detail, we simply choose to use
\begin{gather}\label{eq:vinit}
	\Vinit = \pi b_0 \asp^2 \ \text{(syringe pump)}, \qquad \Vinit = \pi b_0 \Rinit^2 \ \text{(pressurised bottle)},
\end{gather}
which are representative of typical experimental conditions for each injection method. We note that the difference in results between the two cases is due to both the difference between the methods \eqref{eq:flowrate_comp} and the different choices of initial gas volume \eqref{eq:vinit}. The role of these differences and their effect on the two-phase displacement in a rigid cell are investigated in detail in \citet{Cuttle2022}.

\subsection{Non-dimensionalisation}

We non-dimensionalise the governing equations by scaling lengths with the solid thickness $d$, scaling deflections with the initial gap $b_o$, and seeking a balance between all terms in the lubrication equation \eqref{eq:lub}. Thus, the non-dimensional quantities are given by 
\begin{equation}\label{eq:nondim}
  \begin{gathered}
    (\v{x}^*,R^*) = \frac{(\v{x},R)}{d}, \qquad (\v{u}^{s*},w^*, b^*) = \frac{(\v{u}^{s},w, b)}{b_0}, \qquad (p^*,p_b^*, p^{s*},\vv{\sigma}^{s*}) = \frac{(p,p_b,p^s,\vv{\sigma}^s)}{Gb_0/d}, \\
    t^* = \frac{t}{12\mu d^3/(G b_0^3)}, \qquad Q^* = \frac{Q}{2\pi Gb_0^4/12\mu d},
  \end{gathered}
\end{equation}
and the resulting non-dimensional parameters are
\begin{equation}\label{eq:params}
  \begin{gathered}
  \asp^* = \frac{\Rout}{d}, \qquad \fsq^* = \frac{12\mu Q_0 d}{2\pi G b_0^4}, \qquad \fst^* = \frac{d\gamma}{G b_0^2}, \qquad \fca^* = \frac{Gb_0^3}{12d^2\gamma}, \qquad \fht^* = \frac{\pi}{4} \frac{b_0}{d} \fst^*, \\
  \Rinit^* = \frac{\Rinit}{d}, \qquad 	p_a^* = \frac{p_a}{Gb_0/d}, \qquad \Vinit^* = \frac{\Vinit}{2\pi b_0 d^2}.
 \end{gathered}
\end{equation}
Here, $\asp^*$ is the non-dimensional radius of the elastic slab, or equivalently its aspect ratio, and is assumed to be moderately large, while $\fsq^*$ is a non-dimensional flow rate and measures the strength of the fluid--structure interaction in the cell. These two are the main parameters, and also apply to single-phase flow. The three parameters $\fst^*$, $\fca^*$ and $\fht^*$ are related to the role of surface tension, and are in fact related by $\fst^* = (b_0/d)/(12\fca^*) = (d/b_0)(4/\pi)\fht^*$, so only two out of the three are independent.

From here on, we use only non-dimensional quantities, dropping the asterisks for simplicity. The resulting non-dimensional forms of most of the governing equations \eqref{eq:elastic}--\eqref{eq:vinit} are then obtained by simply setting $12\mu = G = b_0 = d = 1$ and replacing $\pi$ in \eqrefsst{eq:injection}{ and }{eq:vinit} by $1/2$. 
The exceptions are the bubble front conditions \eqref{eq:front}, which become
\begin{equation}
  (1 - f_1)\dot R = -b^2\frac{\pd p}{\pd r}, \quad p - p_b = -\frac{2\fst}{b}(1 + f_2) - \frac{\fht}{R} \qquad \text{ at } r = R^+,
\end{equation}
with $f_1$ and $f_2$ functions of $\Ca = \fca \dot R$.

We solve this system numerically using first-order implicit integration in time (backward Euler) and second-order finite differences in space; see \appref{app:num} for details. We typically terminate the simulation when the distance $\asp - R$ from the bubble to the rim decreases below $0.1$, in which case we deem the bubble to be escaping the cell, or when the minimum cell gap 
\begin{gather}
	\bmin(t) = \min_r b(r,t),
\end{gather} 
which typically occurs at a well-defined bulge near the rim, decreases below $0.05$, in which case we deem the cell to be choking, as increasingly fine numerical resolution in space and time would be required to resolve the flow past these thresholds. When the cell is deemed to be choking, increasing the resolution of the simulations indicates that $\bmin$ continues to decrease, and would reach zero in finite time which traps the bubble in the cell, rather than taking infinite time to decay to zero which might allow the bubble to escape. However, the model becomes invalid when the gap is too small; we discuss this issue further in \secref{sec:summ}.

\section{Expansion of the bubble below the choking threshold \label{sec:num}}

Throughout this section, we focus on the specific value $\asp = 20$ for the cell radius and $\fsq = 20$ for the non-dimensional injection flow rate, which is below $\fsq \approx 1.4\asp$ at which the single-phase system is expected to choke \citep{Box2020}. 

\subsection{Review of single-phase flow (no gas)}\label{sec:singlephase}

\begin{figure}
\includegraphics[width=\linewidth]{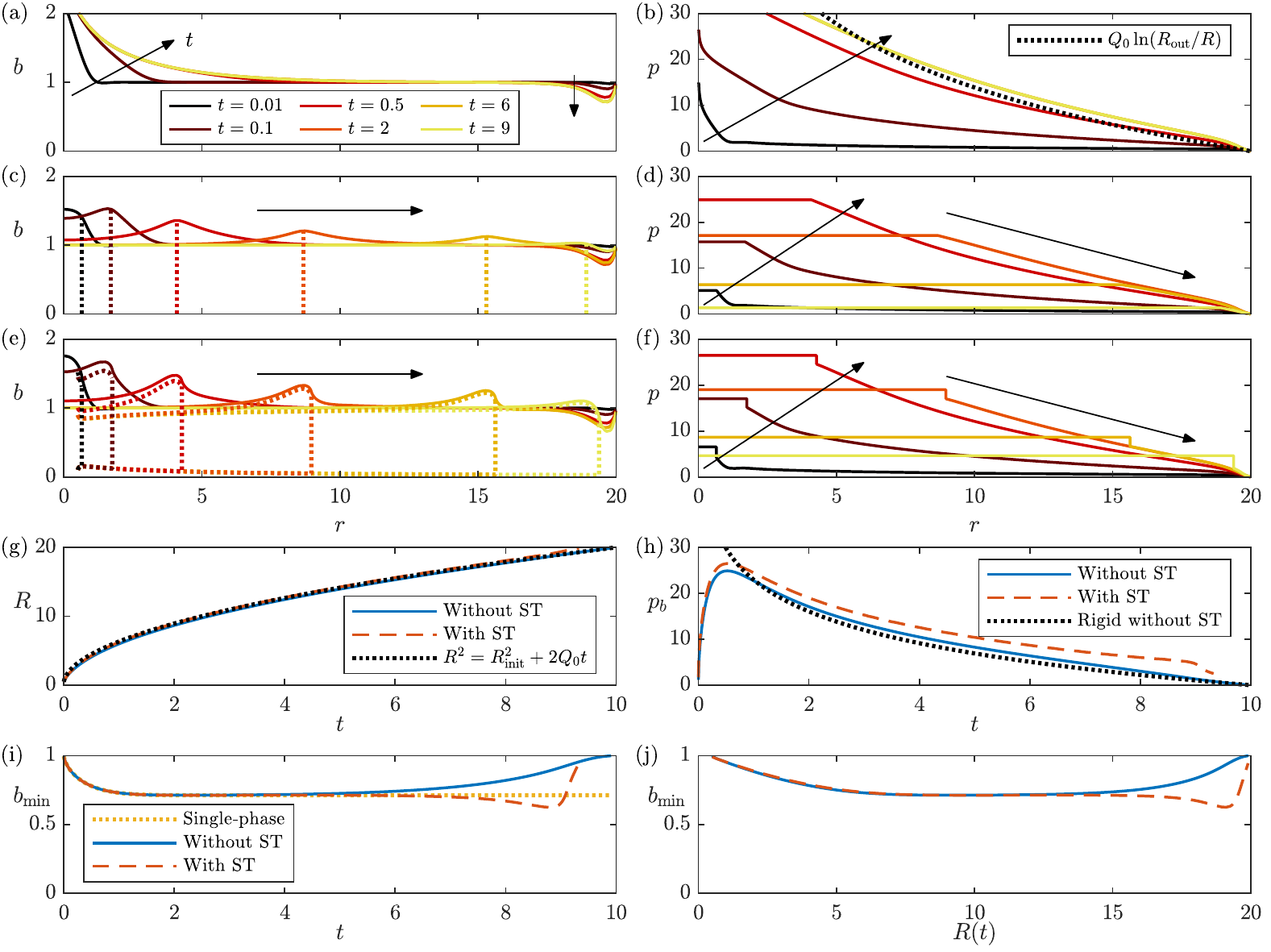}
\caption{Numerical results for incompressible flow with $\fsq = \asp = 20$. Top: Snapshots of channel height/deformation profiles (left) and pressure profiles (right) at various times for \figa{a,b} single-phase flow, \figa{c,d} two-phase flow without surface-tension (ST) effects ($\fst = \fht = \fca = 0$), \figa{e,f} two-phase flow with surface-tension effects ($\fst = 1$, $\fht = 0.1\pi/4$, $\fca = 0.1/12$). In \figa{b}, the pressure profile in a rigid cell (dotted curve) is shown for comparison. In \figa{c}, the vertical dotted lines indicates the position of the displacement front. In \figa{e}, the dotted curves correspond to the bubble boundary, indicating both the position of the displacement front and the thickness of the residual films. Bottom: Time evolution of \figa{g} the bubble radius $R$, \figa{h} the bubble pressure $p_b$ and \figa{i} the minimum cell gap $\bmin$, as well as \figa{j} $\bmin$ plotted against $R$, from \figa{c--f}. In \figa{g,h}, results from a rigid cell with no surface tension are shown for comparison. In \figa{i}, single-phase results from \figa{a,b} are also shown. 
\label{fig:snapshots}}
\end{figure}

We first briefly review the single-phase case, in which there is no gas in the system and flow in the liquid-filled cell is driven by injection of more of the same liquid (so that the lubrication equation \eqref{eq:lub} holds throughout the domain). \Figref[a,b]{fig:snapshots} shows the cell deformation and pressure at various times from a simulation with $\fsq = \asp = 20$. 

We observe that the solid deformation and flow are initially localised near the cell centre (inlet) $r=0$ and the rim (outlet) $r=\asp$. The injected fluid expands the gap near $r=0$ and pushes the solid outward, which in turn bulges near the outlet and squeezes fluid out of the cell at the injection rate. As time passes, the deformation of the solid reaches a steady state, with the pressure profile driving a steady flow through the cell. For a rigid cell, the steady-state pressure profile $p = Q_0 \ln(\asp/R)$ would be reached instantaneously \figparen{dotted curve in \figref[b]{fig:snapshots}}. 

As explained by \citet{Box2020}, the slab deformation is driven by the gradient in normal stress (i.e.~pressure) squeezing the solid towards the rim (rather than by the shear stress from the fluid, which is neglected in this model). Away from the injection point and the rim (i.e.~at distances larger than the solid thickness, $r \gg 1$ and $\asp - r \gg 1$), the solid can be modelled using a long-wave approximation (analogous to fluid lubrication theory) \cite{Box2020, Chandler2021}, which yields the horizontal displacement profile and the surface deflection
\begin{gather}\label{eq:longwave}
	\v{u}_H^s \approx -\frac{1 - z^2}{2}\grad_H p, \qquad w \approx -\grad_H \bcdot \left(\frac{1}{3}\grad_H p\right).
\end{gather}
This explains the somewhat surprising result that there is negligible vertical deflection, $w \ll 1$, for intermediate values of $r$ in \figref[a]{fig:snapshots}, as the harmonic pressure field results in zero vertical deflection and a steady flow. As a result, the steady-state pressure profile in the approximately flat part of the elastic cell differs from that in a rigid cell by an additive constant, corresponding to the additional pressure drop due to the constriction near the rim.

\subsection{Two-phase flow with incompressible gas}\label{sec:incomp}

We now consider the injection of gas. We first neglect any effects of compressibility by imposing the incompressible
injection law \eqref{eq:flowrate_incomp}, and compare the single-phase case discussed previously with a two-phase simulation without surface-tension effects ($\fst = \fht = \fca = 0$) and a two-phase simulation with surface-tension effects ($\fst = 1$, $\fht = 0.1\pi/4$ and $\fca = 0.1/12$ corresponding to a dimensional ratio $b_0/d = 0.1$). 

With no surface tension \figrefp[c,d]{fig:snapshots}, the gap initially expands near the centre and constricts near the rim, as in the single-phase case, and the pressure profiles are similar outside of the bubble region. As the bubble grows outward, the cell relaxes towards its undeformed state behind the advancing bubble front \eqparen{the long-wave approximation \eqref{eq:longwave} for the solid yields $w \approx 0$ for a spatially uniform pressure $p = p_b$}. A localised region of expansion travels with the bubble front, with the solid being squeezed toward the rim on the liquid side while not being squeezed in either direction on the gas side. Near the rim, the bulge initially grows (or equivalently the minimum cell gap $\bmin$ decreases) and then approaches a steady state \figrefp[i]{fig:snapshots}, just like for single-phase flow. However, as the bubble approaches (i.e.~$R \to \asp$), the size of the liquid region (over which the solid is being squeezed towards the rim by the viscous pressure gradient) reduces, and hence the solid starts to relax \figrefp[j]{fig:snapshots}. This is the key mechanism by which the inviscid bubble, due to its proximity to the rim, mitigates the tendency of the system to choke. We will revisit it later in \secref{sec:proximity}. 

The time evolution of the bubble radius $R$ \figrefp[g]{fig:snapshots} closely follows the prediction from a rigid cell, in which conservation of volume yields $R^2 = \Rinit^2 + 2 \fsq t$. This is because the deformation of the soft cell has a relatively small effect on the distribution of the fluids. The bubble pressure $p_b$ \figrefp[h]{fig:snapshots} initially increases as the bulge gap constricts near the rim, but eventually decreases as more and more viscous liquid is replaced by inviscid gas. Due to the constricting bulge near the rim, the pressure remains slightly above the value $p_b = Q_0 \ln(\asp/R)$ it would have in a corresponding rigid-walled cell.

When we include surface tension in the model \figrefp[e,f]{fig:snapshots}, the pressure has a capillary jump at the bubble front (controlled by $\fst$ and $\fht$), which changes the deformation profile in its vicinity. The pressure jumping from a higher value in the bubble to a lower value in the liquid causes the gap to expand immediately behind the bubble front and contract immediately ahead of it \figrefp[e]{fig:snapshots}, as compared with the profile near the interface without surface tension \figrefp[c]{fig:snapshots}. When the bubble approaches the rim, the bulge initially grows slightly due to the pressure jump, before it relaxes due to the reduction in size of the liquid region \figrefp[i,j]{fig:snapshots}. 

The evolution of the bubble radius \figrefp[g]{fig:snapshots} changes slightly due to the change in the cell deformation, and also because of the thin residual liquid films being deposited on the cell walls \figrefp[e]{fig:snapshots}, controlled by the parameter $\fca$. Finally, the bubble pressure \figrefp[h]{fig:snapshots} is larger compared with the simulation without surface tension because of the capillary pressure jump.

\subsection{Two-phase flow with compressible gas}\label{sec:comp}

Next we turn our attention to the effects of the gas compressibility. As can be seen from equations \eqref{eq:flowrate_comp}, compression of the gas simply alters the rate of change of the bubble volume, $Q(t)$, so that it deviates from the nominal value $\fsq$ that is imposed by injection. As a result, the mechanisms for the deformation of the cell, discussed above, remain largely unchanged, but the dynamics of the system may be affected by the varying flow rate $Q(t)$. 

\begin{figure}
\centering
\includegraphics[width=\linewidth]{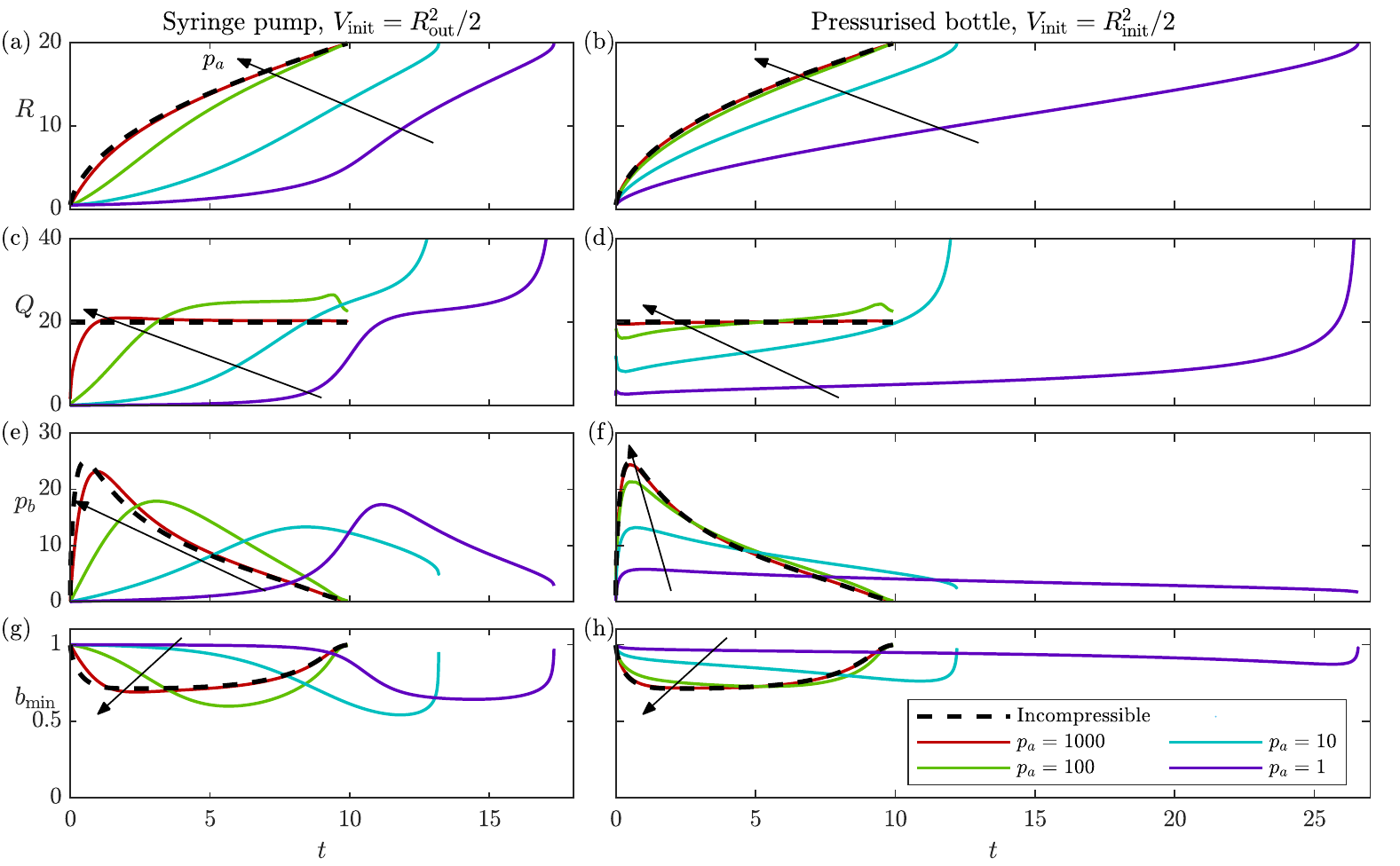}
\caption{Numerical simulations with gas compressibility using the two injection models \eqref{eq:flowrate_comp}, for nominal flow rate $\fsq = \asp = 20$ and four values of the atmospheric pressure parameter $p_a$, without surface tension. Time evolution of \figa{a,b} the bubble radius $R$, \figa{c,d} the liquid flow rate $Q$ exiting the cell, \figa{e,f} the bubble pressure $p_b$, and \figa{g,h} the minimum cell gap $\bmin$. Results from an incompressible simulation \eqref{eq:flowrate_incomp}, which corresponds to $p_a \to \infty$, are shown for comparison. \label{fig:compress}}
\end{figure}

For simplicity, we neglect the effects of surface tension (i.e.~set $\fst = \fht = \fca = 0$), and consider a few different values of the atmospheric pressure $p_a$. The results of our numerical simulations are shown in \figref{fig:compress}, in which we plot the time-evolution of the bubble radius $R(t)$, the rate of change $Q(t)$ of the bubble volume in the cell, the bubble pressure $p_b(t)$ and the minimum cell gap $\bmin(t)$, for injection using either a syringe pump (left column) or a pressurised bottle (right column). We note that, despite the different governing equations \eqref{eq:flowrate_comp} and intial gas volumes \eqref{eq:vinit}, the two injection methods produce qualitatively similar results.

For large $p_a$, which corresponds to the typical gauge pressure in the cell being small compared with atmospheric pressure, the effect of gas compression is negligible: The flow rate $Q(t)$ is approximately equal to the nominal value $Q_0$, and the evolution of the bubble radius $R$, bubble pressure $p_b$ and minimum cell gap $\bmin$ follow the results from the incompressible model. 

As $p_a$ is reduced, the effect of compression becomes significant: The injection initially drives only a small fluid flow $Q(t)$, while the bubble pressure rises and the gas compresses. As the bubble expands and the amount of viscous fluid in the cell reduces, the resistance to flow in the cell decreases, and the bubble attains a maximal pressure before starting to depressurise. However, if there is any remaining overpressure when the bubble reaches the rim, then the flow rate diverges. Also, for larger compressibility, the bubble reaches the rim later. These results are qualitatively similar to those in a rigid cell \citep{Cuttle2022} or a rigid capillary tube \citep{Cuttle2021}. In particular, for a rigid cell with large $\Vinit \gtrsim \Rout^2/2$, the compressibility number defined by \citet{Cuttle2022} is, after the non-dimensionalisation in \eqreft{eq:params}, $C = 4 Q_0 \Vinit/(\asp^2 p_a)$, and is the main parameter that predicts whether the flow rate diverges ($C > 1$) or not ($C \lesssim 1$) as $R \to \asp$ in \figref[c]{fig:compress}. 

The reduction in flow rate due to gas compression initially is the second key mechanism by which the bubble can mitigate the tendency of the system to choke. This will be explored further in \secref{sec:nearchoke}.

\section{The effects of the bubble on choking \label{sec:choke}}

\subsection{The proximity of the bubble to the rim \label{sec:proximity}}

We assess how choking is influenced by the proximity of the inviscid gas bubble to the cell rim by studying the system at a larger flow rate, $\fsq = 29 = 1.45\asp$, that is slightly above the single-phase choking threshold of $\fsq \approx1.4\Rout $ \citep{Box2020}. For simplicity, we once again neglect gas compressibility and surface tension. The evolution of the gap profile near the rim is plotted in \figref[a]{fig:choke}: the bulge grows in amplitude and approaches the opposite wall as the minimum gap $\bmin$ shrinks toward zero. In this case, for which the bubble has initial radius $\Rinit = 0.5$, the bulge develops and the cell chokes before the displacement front is near enough to the rim to have any mitigating effect. Indeed, the profiles agree closely with analogous ones from a single-phase simulation (dashed curves).

\begin{figure}
\centering
\includegraphics[width=\linewidth]{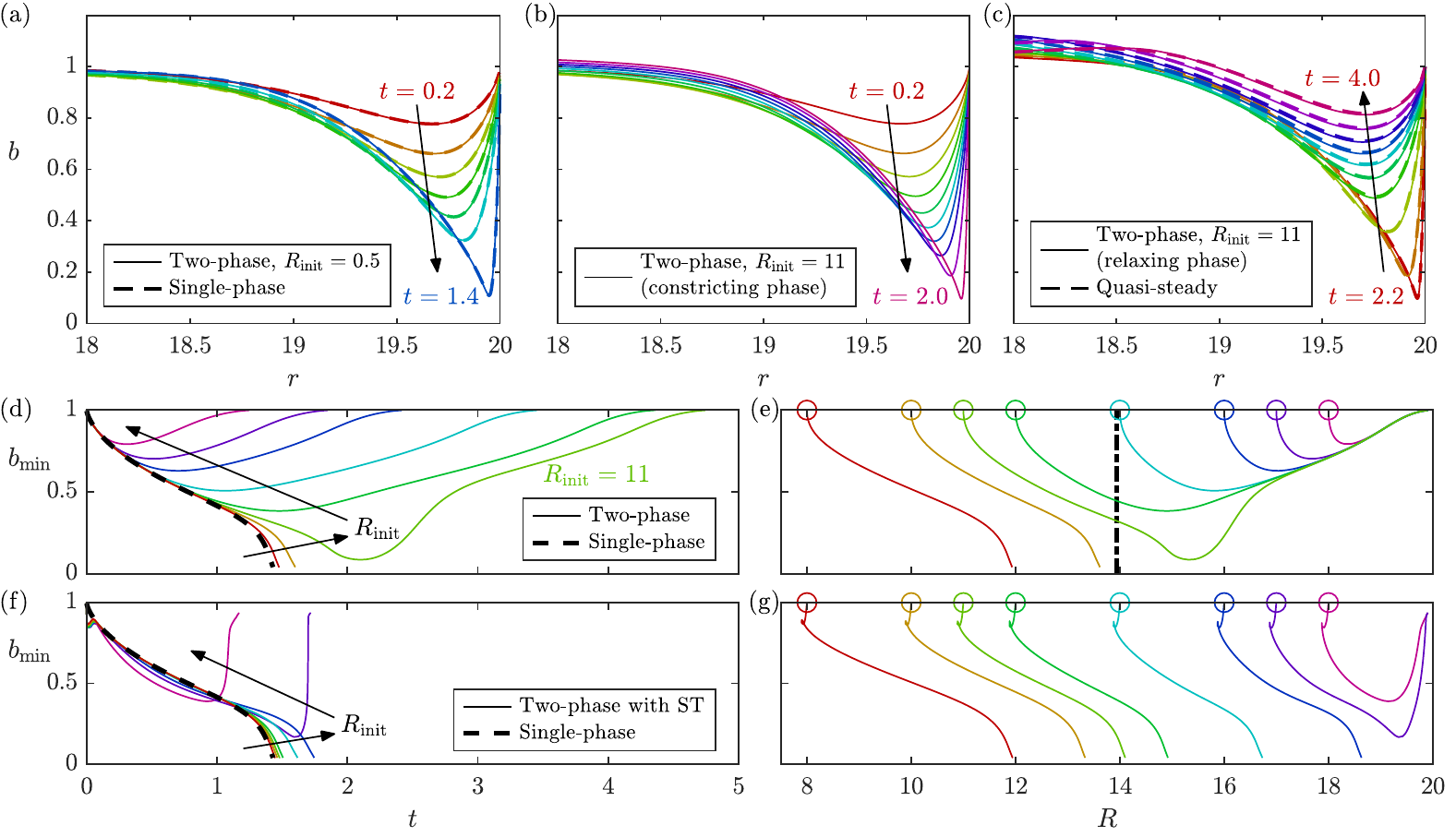}
\caption{Numerical results for incompressible flow with flow rate $\fsq = 29 = 1.45\asp$, which is slightly above the single-phase choking threshold. Top: Snapshots of the channel gap profile $b=b(r, t)$ near the cell rim, in non-dimensional time increments of $0.2$ with first and last times as indicated, without surface tension (ST) and for initial radius \figa{a} $\Rinit = 0.5$ and \figa{b,c} $\Rinit = 11$, split between \figa{b} the constricting phase and \figa{c} the relaxing phase. In \figa{a}, single-phase results are shown for comparison. In \figa{c}, quasi-steady profiles with the same radius and pressure drop are shown for comparison. Bottom: Evolution of the minimum cell gap $\bmin$ as a function of \figa{d,f} time $t$ and of \figa{e,g} the bubble radius $R$ for the two-phase case \figa{d,e} without surface tension and \figa{f,g} with surface tension ($\fst = 1$, $\fht = 0.1 \pi/4$, $\fca = 0.1/12$). Different colours correspond to different initial values $\Rinit = 8,10,11,12,14,16,17,18$, as indicated by the circles in \figa{e,g}, and the curve corresponding to \figa{b,c} is labelled as $\Rinit=11$ in \figa{d}. The single-phase result is also shown in \figa{d,f} for comparison. In \figa{e}, the vertical dash-dotted line shows the choking boundary predicted by the quasi-steady analysis \eqrefp{eq:quasisteady}. \label{fig:choke}}
\end{figure}

Profiles from a simulation with larger initial bubble radius, $\Rinit = 11$, are shown in \figref[b,c]{fig:choke}. The bulge initially grows \figrefp[b]{fig:choke} and the channel nearly chokes. However, as the bubble grows, it reduces the amount of liquid that is squeezing the solid towards the rim. This reopens the channel \figrefp[c]{fig:choke} and choking is averted.

To illustrate how the presence of the bubble near the rim helps the bulge to relax and therefore reduces the tendency of the system to choke, we consider how the bulging changes for different values of the initial bubble radius $\Rinit$. \Figref[d]{fig:choke} shows the time evolution of the minimum cell gap $\bmin$, and the same data is plotted in \Figref[e]{fig:choke} as a function of the interface position $R$. For $\Rinit \lesssim 10$, the bubble does not arrive at the rim early enough to mitigate choking, so $\bmin$ shrinks steadily to zero, reaching it at a finite value of $R<\Rout$. As a result, the system chokes around $t \approx 1.5$, just like in the single-phase case \figparen{dotted curve in \figref[d]{fig:choke}}. For $\Rinit \gtrsim 11$, $\bmin$ initially shrinks, but does not vanish before the bubble is close enough to mitigate the bulge; thereafter, $\bmin$ returns to one instead of decaying to zero. 

\Figref[f,g]{fig:choke} show analogous simulations with surface-tension effects. The capillary pressure drop across the bubble front constricts the gap in front of the bubble (and expands it behind) \figparen{see~\figref[e]{fig:snapshots}} which partly offsets the relaxing effect of the bubble on choking. Hence, the bubble needs to be closer to the rim in order to prevent $\bmin$ decreasing to zero. In these simulations, the system chokes for $\Rinit \lesssim 16$ and only avoids choking for $\Rinit \gtrsim 17$. We do not study the effects of surface tension further.

A further observation that can be made in \figref[e]{fig:choke} is that when the bubble approaches the rim, the curves from different simulations collapse onto a universal curve, indicating that the deformation profile becomes approximately independent of the initial conditions, and instead only depends on the current bubble front position $R$ (as well as the flow rate and the material parameters). We calculate an ad-hoc approximation of this profile by seeking quasi-steady solutions of the governing equations: we neglect the time derivative $\dot{h}$ in the lubrication equation \eqref{eq:lub} and fix the position of the bubble front $R$ instead of evolving it using \eqreft{eq:front}. Thus, we solve the remaining governing equations from \eqrefsst{eq:elastic}{--}{eq:injection} together with
\begin{gather}\label{eq:quasisteady}
	0 = \div \left(b^3\grad p\right) \quad \text{ in } r > R, \quad \text{where $R$ is fixed}.
\end{gather}
The resulting deformation profiles at given values of $R$ \figparen{dotted curves in \figref[c]{fig:choke}} are in excellent agreement with those obtained from the time-evolving simulation, provided that we impose the same bubble pressure $p_b$, rather than the same flow rate $Q$. 

\begin{figure}
\centering
\includegraphics[width=\linewidth]{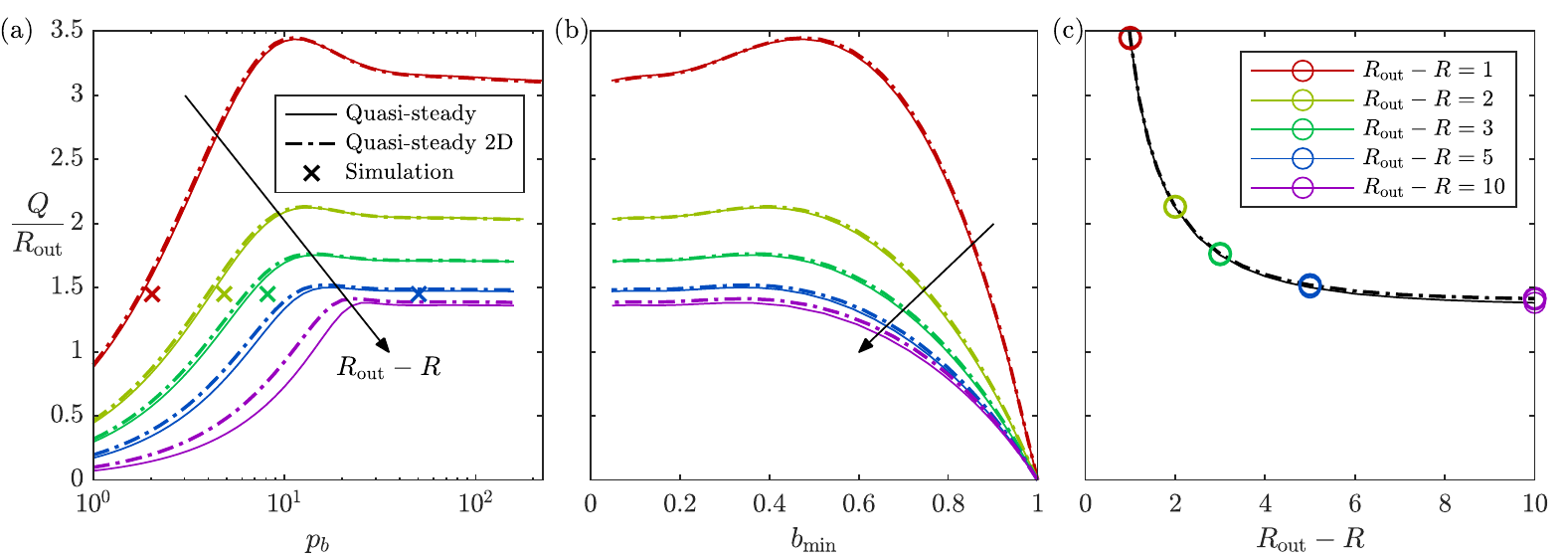}
\caption{Numerical results from the quasi-steady approximation \eqref{eq:quasisteady}, for $\asp = 20$ and without surface tension. The scaled flow rate $Q/\Rout$ is plotted as a function of \figa{a} bubble pressure $p_b$, and \figa{b} minimum cell gap $\bmin$, for various values of the bubble distance from the rim $\asp - R$. The colour-coding is explained with the legend in \figa{c}. \figa{c} The maximal $Q/\Rout$ plotted as a function of $\asp - R$. In all panels, analogous results for a two-dimensional cell are plotted with dash-dotted curves. In \figa{a}, crosses indicate the values of $p_b$ and $Q/\Rout$ at the various distances in the time-evolving simulation from \figref[b,c]{fig:choke}. \label{fig:chokesteady}}
\end{figure}

In this quasi-steady model, any one of $Q$, $p_b$ and $\bmin$ can be treated as the control parameter. We have chosen to performed the quasi-steady calculations for a range of values of the bubble front position $R$ and total pressure drop across the liquid region (or, equivalently, the bubble pressure $p_b$)  rather than $Q$ in order to avoid the issue of multiple solution branches existing for $Q$ just below the maximum value. \Figref{fig:chokesteady} shows how the scaled flow rate $Q/\Rout$ in the quasi-steady solutions depends on $p_b$, $\bmin$ and $R$. For each value of $R$, we see in \figref[a]{fig:chokesteady} that increasing $p_b$ initially drives more flow $Q$, but due to the bulge constricting the channel, $Q$ reaches a maximum and then remains near the maximum as $p_b$ increases further. We also plot the relationship between $\bmin$ and $Q$ \figrefp[b]{fig:chokesteady}, and find similarly that a decrease in $\bmin$ from $1$ initially corresponds to an increase in $Q$, but once the same maximum in $Q$ is reached, the flow rate remains near it as $\bmin$ decreases further. 

We can compare these computations to the results shown in \figref[c]{fig:choke}, in which the flow rate is $\fsq = 1.45\asp$. For each value of $R$ plotted in \figref[a]{fig:chokesteady}, we extract the corresponding values of $p_b$ from the time-evolving simulation in \figref[c]{fig:choke} and mark them with crosses in \figref[a]{fig:chokesteady}. This comparison reveals a small but noticeable difference between the flow rate predicted by the quasi-steady solution and the flow rate obtained in the time-evolving simulations, despite the excellent agreement in deformation profile observed in \figref[c]{fig:choke}.

As described by \citet{Box2020}, in the single-phase case, the occurrence of choking in time-evolving simulations with an imposed flow rate $Q_0$ is linked to the lack of existence of a steady state with flow rate $Q = Q_0$. Analogously, for each value of $R$ we can identify the largest flow rate $Q$ for which a quasi-steady solution exists. The resulting curve \figrefp[c]{fig:chokesteady} represents an approximate boundary, beyond which the large flow rate in the channel is unsustainable and the system is expected to choke. When evolving from an initially undeformed state, which corresponds to $\Rout -R$ decreasing as the bubble grows, the system thus avoids choking if the bubble manages to cross the boundary shown in \figref[c]{fig:chokesteady} before the bulge has had time to grow and make contact with the opposite wall. For example, the boundary for $Q = 1.45$ is at $\Rout - R \approx 6$, i.e.~$R \approx 14$, and indeed as seen in \figref[e]{fig:choke} where this boundary is indicated by the vertical dash-dotted line, in the cases where the cell choked, it did so before the bubble reached $R \approx 14$, while if the bubble did reach $R \approx 14$ then it went on to escape without the cell choking.

The dash-dotted curves in \figref{fig:chokesteady} show the results of the quasi-steady calculations for a two-dimensional cell, in which $Q/\asp$ corresponds to the flow rate per unit length in the third, Cartesian, dimension. The two-dimensional results agree well with the radial results, especially for small $\asp - R$, since the dynamics are limited to the region near the rim where the difference between radial and two-dimensional geometry is small. Hence, the results in \figref{fig:chokesteady} are expected to apply for other cell sizes $\Rout \gg 1$, not just the value $\Rout = 20$ considered here.

\subsection{The near-choking regime for compressible gas \label{sec:nearchoke}}

We now investigate the impact of gas compression on choking, by considering injection with nominal flow rates $Q_0/\asp = 1.5, 2, 3$ and an atmospheric pressure parameter of $p_a = 1000$. To aid the discussion, we plot the flow rate $Q$, bubble pressure $p_b$ and minimum cell gap $\bmin$ as functions of the interface position $R$ in \figref[a--c]{fig:nearchoke}.

\begin{figure}
\centering
\includegraphics[width=\linewidth]{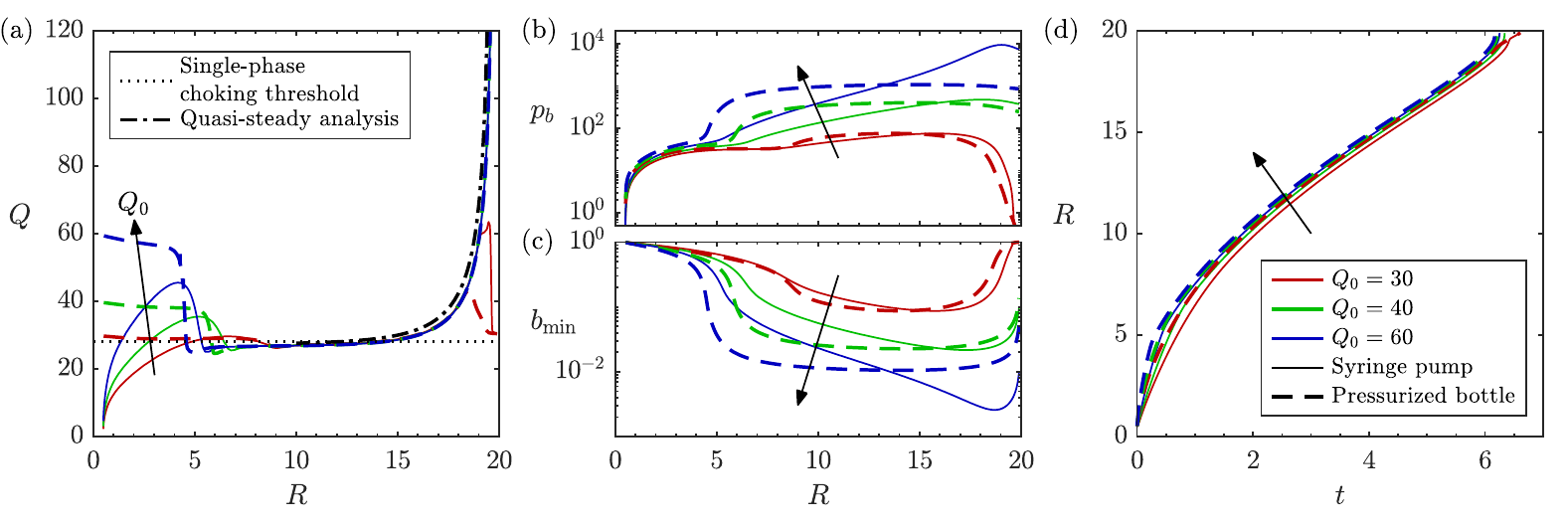}
\caption{Simulations in the near-choking regime, for $\asp = 20$ and three values of the nominal injection flow rate $\fsq$ above the single-phase choking threshold: Evolution of the \figa{a} resulting flow rate $Q$, \figa{b} bubble pressure $p_b$ and \figa{c} minimum cell gap $\bmin$ as functions of the bubble radius $R$, and \figa{d} of $R$ as function of $t$. The simulations assume gas compressibility ($p_a = 1000$) and two injection methods (syringe pump with $\asp^2/2$ or pressurised bottle with $\Vinit = \Rinit^2/2$, shown with solid and dashed curves, respectively), but no surface-tension effects ($\fst = \fht = \fca = 0$). In \figa{a}, the choking threshold for the single-phase flow and results from the quasi-steady analysis \figrefp[c]{fig:chokesteady} are shown for comparison. \label{fig:nearchoke}}
\end{figure}

As evident from \figref[a]{fig:nearchoke}, the flow rate can transiently exceed the critical value for the single-phase flow $Q(t) \approx 1.4\asp$ [horizontal dotted line in \figref[a]{fig:nearchoke}] at early times. However, once the bulge has grown large enough to significantly constrict the gap \figparen{see also \figref[c]{fig:nearchoke}}, the flow rate rapidly drops to this critical value. The mismatch between the larger flow rate $\fsq$ of gas injection and the smaller flow rate $Q(t)$ of liquid exiting the cell is accommodated by volumetric compression of the gas, which causes the gas pressure to increase continually. This in turn reduces the cell gap further \figrefp[b,c]{fig:nearchoke}. Nevertheless, the flow rate does not change significantly, consistent with $Q$ reaching a plateau as $p_b \to \infty$ or $\bmin \to 0$ in the quasi-steady solutions in \figref[a,b]{fig:chokesteady}. As the bubble approaches the rim, the maximum sustainable flow rate increases \figrefp[c]{fig:chokesteady} and the flow rate follows this increase \figparen{dash-dotted curve in \figref[a]{fig:nearchoke}}. \figparen{For $Q_0$ just above the single-phase choking threshold, such as $Q_0=30$ in \figref[a]{fig:nearchoke}, the flow rate stops increasing as the bubble decompresses before escaping the cell, but for larger $Q_0$ the pent up pressure allows the flow rate $Q(t)$, and the bubble velocity $\dot{R}(t)$, to diverge in this model as $R \to \asp$, as discussed at the end of \secref{sec:comp}.}

We conclude that for nominal flow rates above the choking threshold, gas compression enables the system to enter a ``near-choking'' regime after the initial transient. In this regime, the flow rate follows the threshold curve in \figref[c]{fig:chokesteady}, which is a function of the bubble front position $R$, but does not depend on the injection flow rate. A consequence of this is that the simulations with different $Q_0$ and different injection mechanisms all have approximately the same flow rate $Q(t)$ during the main part of the simulation, and hence the time evolution of the bubble radius $R(t)$ is approximately the same between all of them \figrefp[d]{fig:nearchoke}.

We note that compression plays an important role despite the large value of $p_a=1000$ (for which compressive effects were weak in \figref{fig:compress}). Indeed, in the absence of compression, the flow rate $Q(t) = Q_0$ would be sufficiently large for the cell to choke; the minimum cell gap would decrease from its initial value $\bmin = 1$ towards zero, becoming arbitrarily small in finite time \figrefp[d]{fig:choke}. However, as the minimum gap narrows, the viscous resistance (both to flow through the narrow gap and to further reduction of the gap) requires the pressure in the bubble to increase without bound if the flow rate $Q_0$ is to be sustained. As a consequence, no matter how small the compressibility of the gas is, it must compress, which reduces the flow rate $Q(t)$, so that the cell does not choke. Therefore, given that a real gas is never perfectly incompressible, one would expect no choking to occur in experiments. We discuss this apparent contradiction further in \secref{sec:summ}.

\section{The dynamics of the advancing bubble front\label{sec:tw}}

It is possible to elaborate on the dynamics of the advancing bubble front under the assumption that $R \gg 1$ and $\asp - R \gg 1$, i.e.~the bubble and liquid regions have horizontal extents that are large compared with the solid thickness. The elastic equations for the solid in those regions can then be approximated by the long-wave result \eqref{eq:longwave} which yields $w \approx 0$ in both the liquid region \citep{Box2020} and the bubble region. However, the approximation does not apply near the cell rim or near the bubble front, where the horizontal length scale of variation becomes comparable to the solid thickness. Since in this asymptotic regime the bubble is far away from the rim, the deformation near the rim is well described by the single-phase local boundary-layer solution calculated by \citet{Box2020}. Here we study the local behaviour near the bubble front using a travelling-wave approximation.

\subsection{Travelling-wave equations}

We define a local co-moving coordinate $x = r - R(t)$ which is assumed to be $O(1)$. Substituting into the elastic equations \eqref{eq:elastic} and neglecting quantities of order $R^{-1} \ll 1$, we obtain the two-dimensional equations 
\begin{subequations}
\begin{gather}
	\sigma_{xx}^s = -p^s + 2 \pd_x u_x^s, \qquad \sigma_{zz}^s = -p^s + 2 \pd_z u_z^s, \qquad \sigma_{xz}^s = \sigma_{zx}^s = \pd_z u_x^s + \pd_x u_z^s, 
	\\ \pd_x u_x^s + \pd_z u_z^s = 0, \qquad 0 = \pd_x \sigma_{xx}^s + \pd_z \sigma_{zx}^s = \pd_x \sigma_{xz}^s + \pd_z \sigma_{zz}^s.
\end{gather}
\end{subequations}
Under the travelling-wave approximation that the deformation profile is steadily translating with the bubble front $R(t)$, i.e.~$\dot w \approx - \dot{R} w'$, where prime denotes a derivative with respect to $x$, the lubrication equation \eqref{eq:lub} can be integrated to
\begin{gather}\label{eq:tw_lub}
  -\dot R\,w = (1+w)^3 p' + q \quad \text{ in } x > 0,
\end{gather}
where $q$ is a constant of integration. \Eqreft{eq:bub} for the bubble pressure remains as $p = p_b(t)$ in $x < 0$, and the bubble-front conditions \eqref{eq:front} become
\begin{gather}\label{eq:tw_front}
	(1 - f_1(\fca \dot{R}))\dot{R} = -b^2 p', \qquad p - p_b = -\frac{2\fst}{b}(1 + f_2(\fca \dot{R})) \qquad \text{ at } x = 0^+,
\end{gather}
while the conditions on the top and bottom surface of the elastic solid remain as
\begin{gather}
	\v{u}^s|_{z=1} = \v{0}, \qquad w = u_z^s|_{z=0}, \qquad p = -\sigma_{zz}^s|_{z=0}, \qquad 0 = \sigma_{xz}^s|_{z=0}.
\end{gather}
This is a local analysis near $r=R$, so the injection and rim conditions \eqref{eq:injection}--\eqref{eq:flowrate_comp} are irrelevant. Instead we match to the long-wave structure \eqref{eq:longwave} at large $\pm x$, by imposing
\begin{gather}
	w,\ u_z^s \to 0 \text{ as } x \to \pm \infty, 
	\qquad
	u_x^s \to 0 \text{ as } x \to -\infty, 
	\qquad 
	u_x^s \to \frac{1-z^2}{2} q \text{ as } x \to \infty,
\end{gather}
where we identify $q = -\lim_{x \to \infty} p'$ to be the far-field flux or negative pressure gradient. Since the value of $p_b$ simply changes $p$ by a constant, we only need to solve the equations above for $p_b = 0$.

Solving these equations determines the unknown advancement velocity $\dot{R}$ of the bubble, which depends on the non-dimensional surface-tension parameters $\fst$ and $\fca$ and the far-field flux $q$. However, for convenience, we instead proceed by imposing the value of $\dot{R}$ and solving the equations numerically (using Newton iteration) to obtain $q$ as a function of $\dot{R}$. Another important quantity is the effective additional pressure drop in the local region (as compared with an undeformed cell, in which the pressure gradient would be a constant $q$, with no capillary pressure drop),
\begin{equation}
  \Delta p = p_b - \lim_{x\to\infty} \left(p + q\,x\right). 
\end{equation}
This is also calculated numerically as a function of $\dot{R}$.

\subsection{No residual films}

We first consider the case when no residual films are deposited on the walls behind the advancing bubble front, which corresponds to $\fca = 0$. In this case, combining the travelling-wave lubrication equation \eqref{eq:tw_lub} with the kinematic boundary condition \eqref{eq:tw_front} yields the relationship
\begin{gather}\label{eq:twcons_nofilm}
	q = \dot{R},
\end{gather}
meaning that the steady advancement velocity of the bubble must be equal to the depth-averaged lubrication velocity far ahead of the bubble, since liquid volume is conserved.

Channel height profiles for various values of $\dot{R}$ are plotted in \figref[a]{fig:tw} for the case of no surface tension. As was discussed in \secref{sec:incomp}, an advancing bubble is associated with a liquid pressure gradient in $x > 0$ that squeezes the solid away from the bubble and dilates the gap. Results are also included for retreating bubbles ($\dot{R} < 0$), in which case the elastic solid conversely is squeezed towards the bubble and constricts the gap. As $\dot{R}$ decreases towards a critical value just below $-5$, the minimum cell gap shrinks towards zero, and no solutions are found for lower values of $\dot{R}$, indicating an alternative mechanism for choking, in which liquid displacing gas at sufficiently large flow rate causes the elastic solid to make contact with the opposite wall near the moving interface (rather than near the rim of the cell). 

Adding in a static capillary pressure drop $\fst = 1$ across the bubble front \figrefp[b]{fig:tw} results in a relative constriction of the gap ahead of the bubble front and a dilation behind the bubble front, as discussed in \secref{sec:incomp}

The local additional pressure drop in the travelling-wave region is plotted in \figref[c]{fig:tw} as a function of the bubble front velocity $\dot{R}$ for various values of the surface-tension parameter $\fst$. (It is possible to generalise the definition of $\fst$ to include cases of partial wetting with a contact angle $\theta_c$, for which $\fst$ is modified by a factor $\cos\theta_c$ and can therefore be negative.) For the static case $\dot{R} = 0$, the additional pressure drop is simply given by the static formula $\Delta p = 2\Gamma$ (the static deformation profile is an odd function of $x$, so $b=1$ at the bubble front). For $\dot{R} > 0$, the gap expands and the viscous pressure drop reduces, resulting in a smaller $\Delta p$. Similarly, for $\dot{R} < 0$, the gap constricts and the viscous pressure drop increases, but due to the reversed flow direction we again obtain a smaller $\Delta p$.

\begin{figure}
 \includegraphics[width=\linewidth]{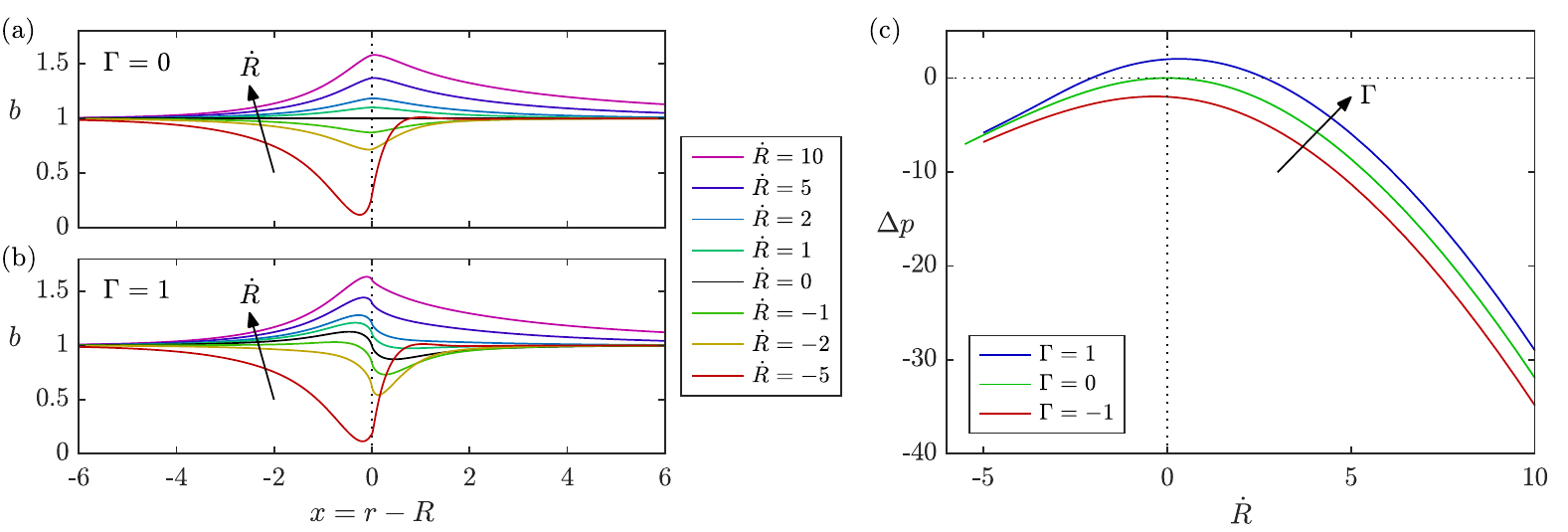}
\caption{Numerically calculated travelling-wave solutions with no residual films ($\fca = 0$). \figa{a,b} Cell gap profiles for various bubble advancement velocities $\dot{R}$ and two values of the surface-tension parameter $\fst$. The location $x=0$ of the bubble front is indicated with a vertical dotted line. \figa{c} The local additional pressure drop as a function of velocity $\dot{R}$ for three values of $\fst$. \label{fig:tw}}
\end{figure}

\subsection{With residual films}

We now consider the case of non-zero $\fca$, representing the deposition of thin liquid films on the cell walls behind the advancing bubble front. Combining \eqreft{eq:tw_lub} with \eqreft{eq:tw_front} now yields a more complicated relationship between the far-field flux $q$ and the advancement velocity $\dot{R}$, which we can express in terms of the total thickness $m$ of films deposited on the walls as
\begin{gather}\label{eq:twcons_film}
	q = (1-m)\dot{R}, \qquad m = f_1(\fca \dot{R}) b|_{x=0}.
\end{gather}

\begin{figure}
\includegraphics[width=\linewidth]{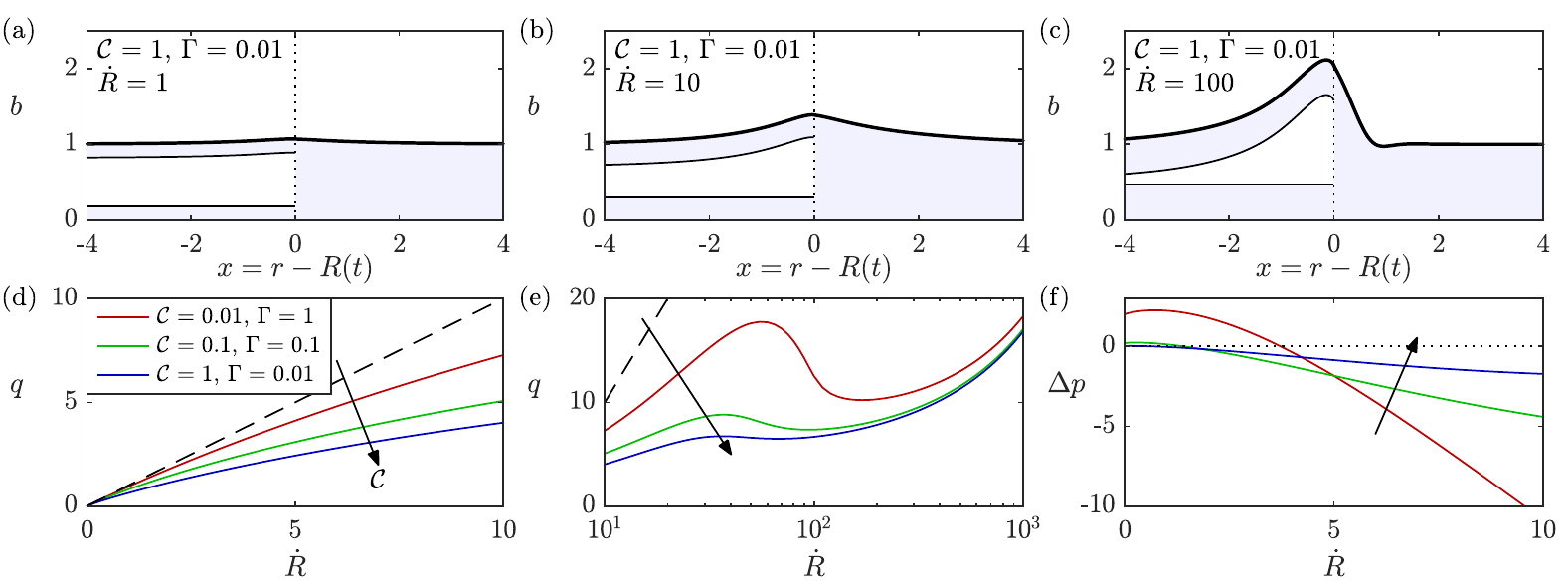}
\caption{Numerically calculated travelling-wave solutions for an advancing bubble front that leaves behind liquid films on the cell walls. Top row: Deformation profiles (thick lines) and residual film thicknesses (thin lines) for $\fca = 1$, $\fst = 0.01$, and three values of the front velocity: \figa{a} $\dot{R}=1$, \figa{b} $\dot{R}=10$ and \figa{c} $\dot{R}=100$. Bottom row: The \figa{d,e} far-field flux $q$ and \figa{f} local additional pressure drop $\Delta p$ as functions of $\dot{R}$ for three different values of $(\fca,\fst)$ corresponding to a dimensional ratio $b_0/d = 0.12$. In \figa{d,e}, the result \eqref{eq:twcons_nofilm} for $\fca = 0$ (and any value of $\fst$) is shown (dashed line) for reference. \label{fig:twfilms}}
\end{figure}

Examples of resulting channel height profiles are plotted in \figref[a--c]{fig:twfilms}, with the thin curves in the bubble region $x < 0$ showing the residual liquid films of thickness $m/2$ coating each wall. As $\dot{R}$ increases, both the film correction factor $f_1(\fca \dot{R})$ and the cell gap $b|_{x=0}$ increase, which results in the residual film thickness $m$ increasing and the ratio $q/\dot{R}$ decreasing. For small and moderately large $\dot{R}$, for which $m$ is not too close to $1$, the bubble continues to push a significant amount of liquid ahead of it, with the far-field flux being $q = O(\dot{R})$ \figrefp[d]{fig:twfilms}. However, as $m$ approaches $1$, the bubble transitions to ``peeling'' the two walls apart while leaving the fluid mostly in place as two thick films coating the walls \figrefp[c]{fig:twfilms}. This allows the advancement velocity to become much larger than the far-field flux \figrefp[e]{fig:twfilms}. (In practice, for large $\dot{R}$, rather than settling into a steadily translating state, the system might exhibit unsteady dynamics such as repeated pinch-off of bubbles as the residual films make contact and reconnect, and become more susceptible to instability in the third dimension.)

In \figref[f]{fig:twfilms} we plot the local additional pressure drop as a function of $\dot{R}$. For the same value of $\dot{R}$, the flow rate $q$ is lower \eqrefp{eq:twcons_film} compared with the case without films \eqrefp{eq:twcons_nofilm}, and hence the effect of the deformation of the cell on the pressure drop is also reduced. Therefore, for the same value of $\dot{R}$, the magnitude of the local additional pressure drop can be significantly smaller in the case with films compared to the case without films \figparen{compare \figref[f]{fig:twfilms} with \figref[c]{fig:tw} at, e.g., $\dot{R}=10$}.

\subsection{Comparison with numerical simulations}

In order to apply the travelling-wave analysis to the time-evolving problem, we combine it with the long-wave approximation in the liquid region and the local single-phase bulge solution near the rim. In the long-wave liquid region, we have $w=0$ and hence, by conservation of volume, $\nabla_H^2 p = 0$. This results in
\begin{gather}
  p = Q \ln \frac{\asp}{r} + \Delta p_\bulge(Q/\asp),
\end{gather}
where the additional pressure drop $\Delta p_\bulge$ near the rim due to the bulging is a function of the local flux $Q/\asp$ and can be extracted from the local two-dimensional solutions of \citet{Box2020}. From this, we can deduce the value of the matching quantity $q = q_\front(\dot R)$, and express the bubble pressure in terms of the matching quantity $\Delta p = \Delta p_\front(\dot R)$,
\begin{gather}\label{eq:tw_app}
	q_\front(\dot R) = \frac{Q}{R}, \qquad p_b = Q \ln \left(\frac{\asp}{R} \right) + \Delta p_\bulge(Q/\asp) + \Delta p_\front(\dot R).
\end{gather}
For imposed $Q = Q_0$ and a known initial value of $R$, the first equation in \eqreft{eq:tw_app} can be integrated numerically to yield the evolution of $R$. For the case of compressible gas injection \eqrefsp{eq:flowrate_comp}, or other methods of injection that depend on $p_b$, the evolution of $R$ is obtained by solving \eqreft{eq:tw_app} coupled to the injection condition.

\begin{figure}
  \includegraphics[width=\linewidth]{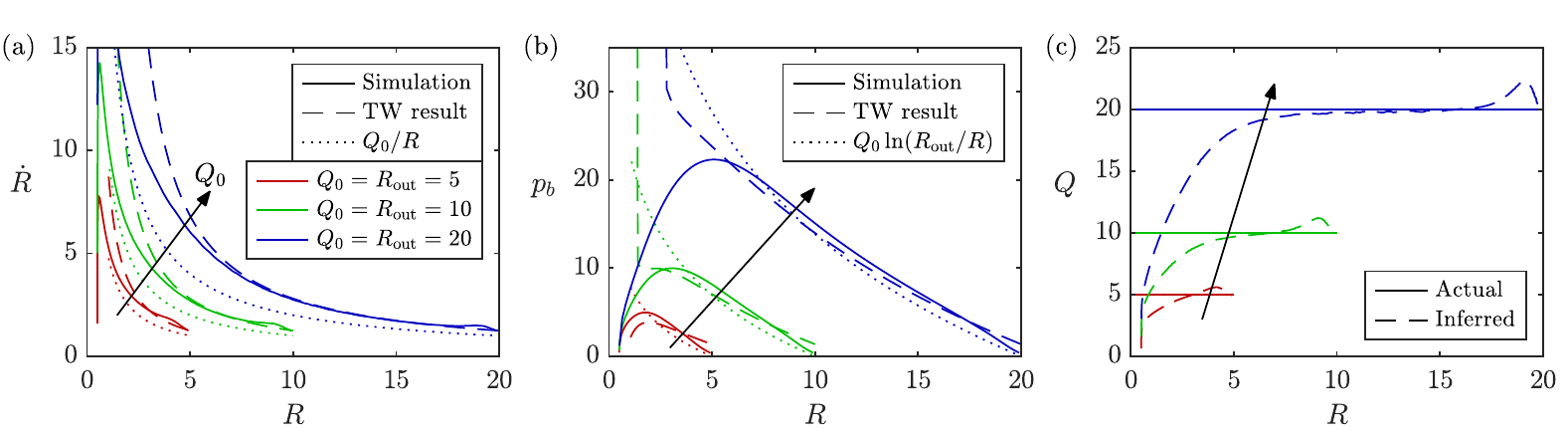}
\caption{Application of travelling-wave results to the time-evolving problem, with three values of $\fsq = \asp$, no gas compression and $\fst = 0.1$, $\fht = 0.01\pi/4$, $\fca = 1/12$. The evolution of the bubble \figa{a} front velocity $\dot{R}$ and \figa{b} pressure $p_b$, as a function of its position $R$, obtained in the simulations and using the travelling-wave approximation \eqref{eq:tw_app}. The results $\dot{R} = Q_0/R$ and $p_b = Q_0 \ln(\Rout/R)$ for a bubble without surface tension in a rigid cell are also shown. \figa{c} The imposed flow rate of liquid exiting the cell and the flow rate inferred by applying the travelling-wave analysis to the $R=R(t)$ data from the simulations. \label{fig:twapp}}
\end{figure}

We compare results from the travelling-wave analysis with results from a time-evolving simulation, focusing on a case with no gas compression, moderate effects of surface tension and thin films ($\fst = 0.1$, $\fht = 0.01 \pi/4$ and $\fca = 1/12$, corresponding to a dimensional ratio $b_0/d = 0.1$), and three different values of the cell radius, $\asp = 5,10,20$ \figrefp{fig:twapp}{}. \Figref[a]{fig:twapp} shows the velocity $\dot{R}$ of the bubble front, as a function of its position $R$, comparing the values obtained in the simulations (solid curves) to the predictions from the travelling-wave analysis (dashed curves). As expected, there is good agreement between the two for $\asp = 20$ and intermediate values of $R$, when the bubble front is far away from the centre and the rim of the cell so that the long-wave approximation holds. For the smaller values of $\asp$, the bubble cannot be as far away from both regions of the cell, so the agreement is worse. For comparison, the dotted curves show the prediction without residual films \eqrefp{eq:twcons_nofilm}, i.e.~$q_\front(\dot{R}) = \dot{R}$, which is noticeably different. \Figref[b]{fig:twapp} shows the bubble pressure $p_b$ as a function of $R$. Once again, the agreement between simulations and predictions is the best for large $\asp$.

The travelling-wave results can also be used to infer the flow rate $Q = q_\front(\dot R)\,R$ from the evolution of $R(t)$. \Figref[c]{fig:twapp} shows the results (dashed curves) when applied to the data from the simulations in \figref[a,b]{fig:twapp}. The best agreement with the true value of $Q$ (solid lines) is obtained for large $\asp$, as expected.

\section{Discussion \label{sec:summ}}

We have presented and analysed an axisymmetric model for injection of a gas bubble into a liquid-filled elastic-walled Hele-Shaw cell bounded by a confined incompressible elastic solid \figrefp{fig:setup}{}. For injection of the same viscous liquid rather than gas, the cell is known to choke for flow rates exceeding a critical value. This choking occurs because the pressure gradient of the viscous flow squeezes the elastic solid towards the rim, where it bulges into the channel and makes contact with the opposite wall \citep{Box2020}. 

We have identified two mechanisms by which injection of a gas bubble instead of viscous liquid reduces the tendency of the cell to choke. Firstly, for a given flow rate, the proximity of the inviscid bubble to the cell rim reduces the size of the liquid region over which the cell is being squeezed towards the rim by the viscous pressure gradient. Using a quasi-steady analysis, we have obtained an approximation for the increased choking threshold as a function of the distance from the bubble to the rim \figrefp[c]{fig:chokesteady}. (The surface tension of the bubble can counteract this effect slightly, due to the capillary pressure drop causing a constriction of the cell ahead of the bubble, which warrants further investigation.) Secondly, compression of the gas reduces the flow rate of the liquid, and since choking requires the pressure to diverge, choking with a compressible gas is not possible. Instead, the gas compresses to keep the flow rate below the choking threshold, resulting in a near-choking behaviour in which the liquid flow rate closely follows the bubble-position-dependent theoretical threshold regardless of the nominal injection rate of the gas \figrefp[a]{fig:nearchoke}. 

The near-choking regime is similar to phenomena observed in other fluid--structure interaction (FSI) problems. For example, when a fluid is driven through a confined, deformable porous medium, the imposed pressure gradient squashes the medium against the outlet, which reduces the permeability and ultimately restricts the outflow, i.e.~the fluid flux reaches an upper bound and becomes insensitive to further changes of the pressure head~\citep{Hewitt16}. Flow saturation also occurs for inertial flow of a viscous fluid in finite-length elastic tubes: the increasing pressure head reduces the cross-sectional area of the tube leading to increase in the local fluid velocity, which in turn reduces the internal fluid pressure via the Bernoulli effect and causes further constriction of the tube ~\citep{Heil03}. Inherently, all of these mechanisms rely on interactions between a flow and an elastic structure, though the details of the FSI are different to the ones considered here.

The study of choking involves the cell gap shrinking to zero. However, our model is formally not valid once the gap becomes too small, as other effects become important, such as adhesion forces between the walls, small-scale roughness of the surfaces, deviations from perfect axisymmetry, and (eventually) the breakdown of the continuum approximation. All of these effects are likely to promote choking by locally enabling initial contact between the walls in isolated azimuthal regions without incurring a divergent pressure. Hence, for example in the near-choking regime, although our model always predicts a very small but non-zero cell gap, in actuality the walls can make contact with each other and choke the flow. This presumably also explains why choking is readily observed in the experiments of \citet{Box2020} and \citet{Peng2022}.

When the radius of the elastic solid is very large compared with its thickness, long-wave approximations can be applied in the bubble and liquid regions. We have shown that in this regime the elastic cell behaves like a rigid cell, but with modified kinematic and dynamic conditions at the advancing bubble front due to the deformation near the front, and a modified outlet pressure condition due to the bulging near the rim (\secref{sec:tw}). Although we have assumed axisymmetry in the present study, these approximations readily extend to non-axisymmetric flows. As a result, the viscous-fingering instability in the elastic-walled cell can be simulated using a standard Hele-Shaw solver for a rigid cell but with modified boundary conditions. Another application for the modified kinematic conditions at the bubble front is to infer the local flux, and hence the global flow rate, from non-axisymmetric experimental data for the position of the bubble front, as was done by \citet{Peng2022}.

In their experiments performed at larger values of $Q_0$, \citet{Peng2022} suggested that compressibility of the elastic solid will begin to play a role in the problem as the injection pressure approaches a non-negligible fraction of the bulk modulus of the elastomer. It is straightforward to adapt the present model to account for solid compression, which introduces another non-dimensional parameter in the form of Poisson's ratio $\nu$. However, analysis of the model becomes more difficult, as the long-wave approximation is significantly more complicated \citep{Chandler2021} and the travelling-wave solutions depend on both $\nu$ and the bubble pressure $p_b$.

\begin{acknowledgments}
The authors thank F.~Box and A.~Juel for discussions. The work of the group from Manchester was funded by the EPSRC [Grant No.~EP/R045364/1]. The work of the group from Oxford was funded by the ERC under the European Union's Horizon 2020 Programme [Grant No.~805469] and by the EPSRC [Grant No.~EP/P009751/1]. 
\end{acknowledgments}

\appendix

\section{Numerical method \label{app:num}}

We have implemented a finite-difference scheme in Matlab, making use of its built-in routines for LU factorisation and sparse matrix solution. The solid domain $0 \leq r \leq \asp$, $0 \leq z \leq 1$ is discretised using a grid with an initially uniform spacing of $0.02$. The radial grid is adapted non-uniformly as required to keep the grid spacing below $2\%$ of the estimated local length scale, and the vertical grid is also refined near the surface to keep the smallest grid cells nearly square. 

The solid displacements $u_r^s$ and $u_z^s$ are measured at the midpoint of the horizontal and vertical cell boundaries, respectively, and the solid pressure $p^s$ is measured at the midpoint of each cell. The associated equations for $u_r^s$, $u_z^s$ and $p^s$ are
\begin{equation}
  -\pd_r p^s + \left[\pd_r^2 + \tfrac1r \pd_r - \tfrac{1}{r^2} + \pd_z^2\right]u_r^s = 0, \qquad
	-\pd_z p^s + \left[\pd_r^2 + \tfrac1r \pd_r + \pd_z^2\right] u_z^s = 0, \qquad
	\left[\pd_r + \tfrac{1}{r}\right]u_r^s + \pd_z u_z^s = 0,
\end{equation}
which are evaluated using standard second-order finite differences. The singularity in the solid equations at $(r,z)=(R,0)$, due to the discontinuity $\Delta p$ in the cell pressure $p$ at $r = R$, is treated analytically in a small neighbourhood of the bubble front by subtracting a two-dimensional leading-order solution,
\begin{gather}
u_x^s = -\Delta p\,\frac{z(1+\ln(x^2+z^2))}{4\pi}, \qquad
u_z^s = -\Delta p\,\frac{x(1-\ln(x^2+z^2))}{4\pi}, \qquad
p^s = \Delta p\,\frac{\pi - 2\arctan(x/z)}{2\pi},
\end{gather}
where $x = r - R$, which yields additional terms that are proportional to $\Delta p$ in the equations.

We define an integrated surface displacement $\psi$, measured on the radial cell boundaries, with the associated equation $\psi = \int_r^{\asp} wr\,\mathrm{d}r$. This allows the equations \eqref{eq:lub} for the gas and liquid to be written as
\begin{equation}
  p = p_b \quad \text{ in } r < R, \qquad r\,\pd_r p = -\frac{Q + \dot{\psi}}{b^3} \quad \text{ in } R < r < \asp,
\end{equation}
which we take to be the equations associated with the variable $p$. 

The time derivative is discretised implicitly as $\dot{\psi} = (\psi - \psi|_{\subtext{prev}})/\Delta t$, and $\dot{R} = (R - R_{\subtext{prev}})/\Delta t$, in which $\psi|_{\subtext{prev}}$ and $R_{\subtext{prev}}$ are the known values from the previous time step, while all other unknowns are to be determined at the current time step. The temporal step size $\Delta t$ is adapted to keep the temporal resolution around $0.5\%$. The resulting large non-linear system of equations is solved using Newton iteration (using the previous values as starting guess), with a decomposition into linear and non-linear parts to increase efficiency, as follows.

We collect the values of $u_r^s$, $u_z^s$, $p^s$ and $\psi$ in a solution vector $\v{X}_L$, while the values of $p$ and other individual quantities such as $R$, $\dot{R}$, $Q$, $p_b$, $b|_{r=R}$ and $\Delta p$ are collected in $\v{X}_N$. The complete set of discretised equations to be solved can then be represented as $\v{F}_L(\v{X}_L,\v{X}_N) = \v{0}$ and $\v{F}_N(\v{X}_L,\v{X}_N) = \v{0}$, for the equations associated with $\v{X}_L$ and $\v{X}_N$, respectively. Given a guess $(\v{X}_L,\v{X}_N)_i$ for the solution vectors, the residuals $\v{F}_{L,N}$ and the Hessian are calculated, and an equation 
\begin{equation}
  \begin{pmatrix} \v{F}_L \\ \v{F}_N \end{pmatrix} + 
	\begin{pmatrix} A_{LL} & A_{LN} \\ A_{NL} & A_{NN} \end{pmatrix}
	\left[\begin{pmatrix} \v{X}_L \\ \v{X}_N\end{pmatrix}_{i+1} - \begin{pmatrix} \v{X}_L \\ \v{X}_N\end{pmatrix}_{i}\right] = \v{0}
\end{equation}
for the next iteration is obtained. Here, due to the decomposition into $L$ and $N$ parts, the largest matrix, $A_{LL}$, is a constant, so its LU factorisation can be precomputed and stored (every time the grid is altered), which allows the product of $A_{LL}^{-1}$ with vectors to be calculated efficiently. We then eliminate $\v{X}_L$ from the equations and obtain an expression for $\v{X}_N$ which requires no matrix inversions apart from the precomputed $A_{LL}^{-1}$ and the solution of a matrix equation of approximate size $N_r$ (the number of radial grid points). Although this matrix equation is dense, it is much faster to solve than the original sparse matrix equation of approximate size $3N_r N_z$ (where $N_z$ is the number of vertical grid points).

\bibliography{manuscript}

\end{document}